\documentclass[cits]{PoS}

\usepackage{amsmath}
\usepackage{amssymb}
\usepackage{graphicx}
\usepackage{dsfont}
\usepackage{dcolumn}
\usepackage{units}
\usepackage{bm}
\usepackage{wasysym}
\usepackage{multirow}
\usepackage{slashed}

      \newcommand{\be}{\begin{equation}}
      \newcommand{\ee}{\end{equation}}

      \definecolor{violet}{RGB}{111,0,255}
      \definecolor{webgreen}{rgb}{0,0.75,0}
      \definecolor{webred}{rgb}{0.75,0,0}
      \definecolor{webblue}{rgb}{0,0,0.75}
      \definecolor{darkblue}{rgb}{0,0,0.6}
      \definecolor{darkgreen}{rgb}{0,0.5,0.5}
      \definecolor{darkpurple}{rgb}{0.5,0,0.5}
      \definecolor{darkorange}{rgb}{1,0.5,0}
      \definecolor{darkgrey}{rgb}{0.4,0.4,0.4}
      \definecolor{lgray}{rgb}{0.95,0.95,0.95}
      \definecolor{lgreen}{rgb}{0.95,1.00,0.90}
      \definecolor{lred}{rgb}{1.00,0.90,0.80}
      \definecolor{lblue}{rgb}{0.2,0.35,1.00}
      \definecolor{shadecolor}{rgb}{1.00,0.92,0.82}

\title{Towards resonance properties in the Dyson-Schwinger approach}

\ShortTitle{Resonances in the Dyson-Schwinger approach}

\author{Gernot Eichmann\thanks{Supported by the Portuguese Science Foundation under 
      FCT Investigator Grant IF/00898/2015.}\\
        Instituto Superior T\'ecnico, Universidade de Lisboa, 1049-001 Lisboa, Portugal \\
        E-mail: \email{gernot.eichmann@tecnico.ulisboa.pt}}

\abstract{We give a brief summary of the Dyson-Schwinger and Bethe-Salpeter approach to hadron spectroscopy and 
          report on recent progress in determining resonance properties in this framework.
          We exemplify the extraction of resonances using a scalar model, where we solve the scattering equation
          for the four-body scattering amplitude and extract the pole locations on the second Riemann sheet as well as the phase shifts.}

\FullConference{Light Cone 2019 - QCD on the light cone: from hadrons to heavy ions - LC2019\\
		16-20 September 2019\\
		Ecole Polytechnique, Palaiseau, France}

\begin{document}

     \begin{figure*}[t!]
     \center{
     \includegraphics[width=0.9\textwidth]{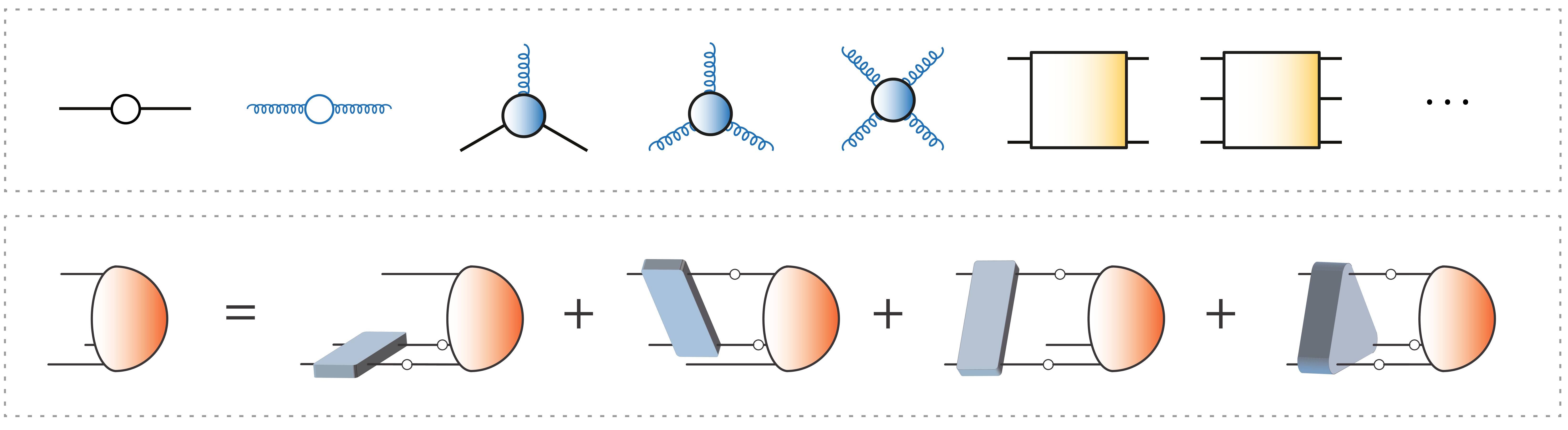}}
        \caption{QCD's elementary $n$-point functions \textit{(top)} and three-body Bethe-Salpeter equation \textit{(bottom)}.}
        \label{fig:nptfcts}
     \end{figure*}

     Understanding the properties of resonances is a central task in nonperturbative QCD studies.
     Most of the `bound states' in QCD are in fact resonances which decay and thus correspond to poles in the complex momentum plane of scattering amplitudes.
     In the following we discuss resonances in the context of the functional approach of Dyson-Schwinger equations (DSEs) and Bethe-Salpeter equations (BSEs), 
     where progress has been made in the calculation of hadron spectra and matrix elements; 
     see e.g. the reviews~\cite{Cloet:2013jya,Eichmann:2016yit,Sanchis-Alepuz:2017jjd} and references therein.
     
     The basic idea of the DSE/BSE approach is visualized in Fig.~\ref{fig:nptfcts}. 
     The starting point are QCD's correlation functions or $n$-point functions shown in the upper panel: 
     the dressed quark and gluon propagators, the quark-gluon vertex, three-gluon vertex etc., which
     encode the properties of the quantum field theory.
     Their momentum dependence and Lorentz/Dirac structure can be calculated using nonperturbative methods such as
     (gauge-fixed) lattice QCD~\cite{Cucchieri:2011ig,Maas:2011se,Aouane:2012bk,Sternbeck:2017ntv,Boucaud:2017obn,Oliveira:2018lln,Sternbeck:2019twy},  
     the functional renormalization group (FRG)~\cite{Pawlowski:2005xe,Mitter:2014wpa,Cyrol:2016tym,Cyrol:2017ewj,Alkofer:2018guy} or the DSEs 
     which are the quantum equations of motion~\cite{Roberts:1994dr,Alkofer:2000wg,Fischer:2006ub}.
     An agreement between these methods has emerged over the last years at the level of two- and three-point correlation functions,
     whereas the structure of higher $n$-point functions is less well known but under active investigation; see~\cite{Cyrol:2016tym,Cyrol:2017ewj,Eichmann:2014xya,Williams:2014iea,Cyrol:2014kca,Binosi:2014kka,Aguilar:2015bud,Eichmann:2015nra,Huber:2018ned} and references therein.
     
     Hadron properties are contained in $n$-point functions that admit a color-singlet spectral representation. In Fig.~\ref{fig:nptfcts},
     the quark-antiquark four-point function contains meson poles, the six-point function baryon poles, etc.
     A stable hadron corresponds to a pole on the real axis and a resonance to a pole in the complex plane of a higher Riemann sheet.
     This is also what allows one to derive a BSE at the respective pole; for example, the three-body BSE (or covariant Faddeev equation)
     in the lower panel of Fig.~\ref{fig:nptfcts} describes the binding of three valence quarks to a baryon, where the kernels again depend on QCD's $n$-point functions.
     Similar relations allow one to calculate matrix elements such as form factors and scattering amplitudes.

      \begin{figure*}[t]
              \centering
              \includegraphics[width=\textwidth]{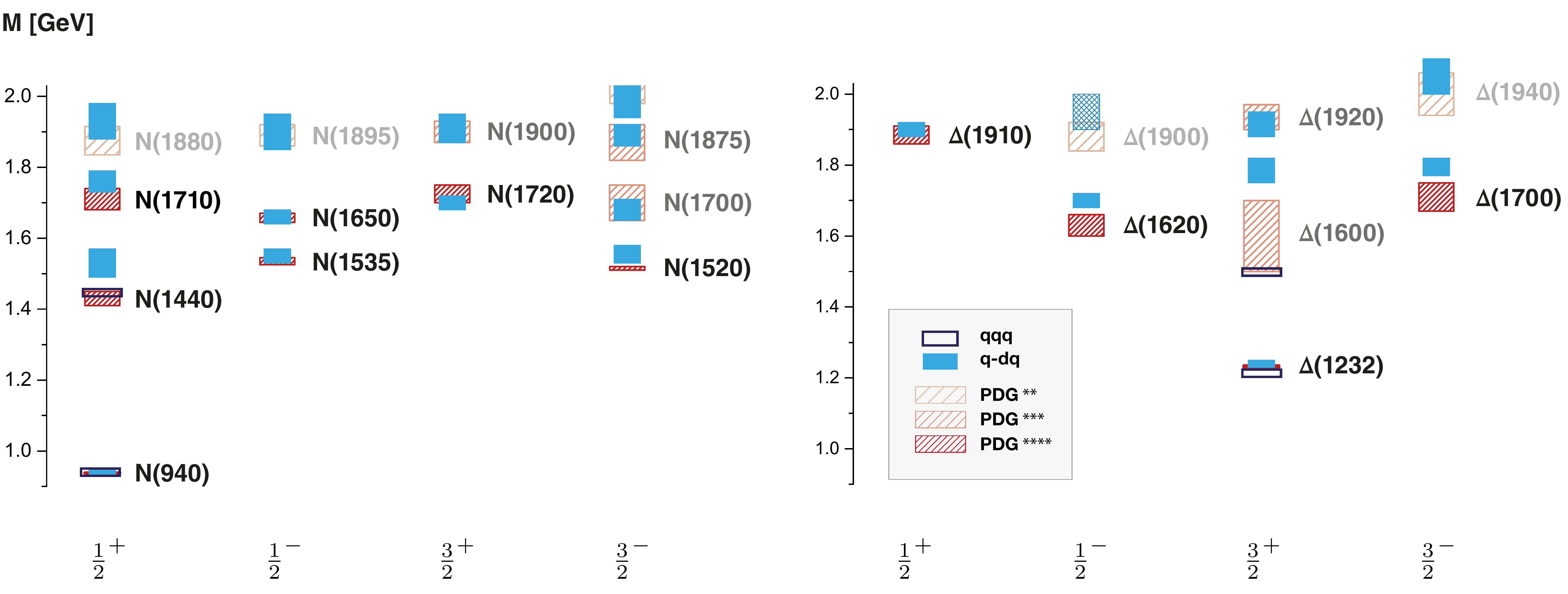}
              \caption{Nucleon and $\Delta$ spectrum for $J^P=1/2^\pm$ and $3/2^\pm$ states.
                       The three-body (open boxes~\cite{Eichmann:2009qa,SanchisAlepuz:2011jn})
                       and quark-diquark results (filled boxes~\cite{Eichmann:2016hgl})
                       are compared to the PDG values with their experimental uncertainties~\cite{Tanabashi:2018oca},
                       see \cite{Eichmann:2016hgl} for details.
                       } \label{spectrum-light}
      \end{figure*}
     
     Even though our knowledge of the underlying $n$-point functions has considerably improved in recent years,
     their implementation in BSEs is not straightforward due to
     chiral symmetry constraints, see e.g.~\cite{Chang:2009zb,Williams:2015cvx}.
     These ensure that the pseudoscalar mesons become massless in the chiral limit, while
      the quark DSE dynamically breaks
      chiral symmetry and generates a running quark mass function.
     The starting point for systematic kernel constructions is the rainbow-ladder truncation, which approximates
      the two-body kernels by effective gluon exchanges~\cite{Maris:1997tm,Maris:1999nt} and
      describes a range of meson and baryon properties including masses, 
      decays, elastic and transition form factors well~\cite{Eichmann:2016yit}.
      Results for the light baryon spectrum, both from the three-body BSE and its quark-diquark simplification, are shown in Fig.~\ref{spectrum-light};
      excited hyperons are discussed in~\cite{Chen:2017pse,Fischer:2017cte,Eichmann:2018adq,Yin:2019bxe}.

\begin{figure*}[b]
              \centering
              \includegraphics[width=\textwidth]{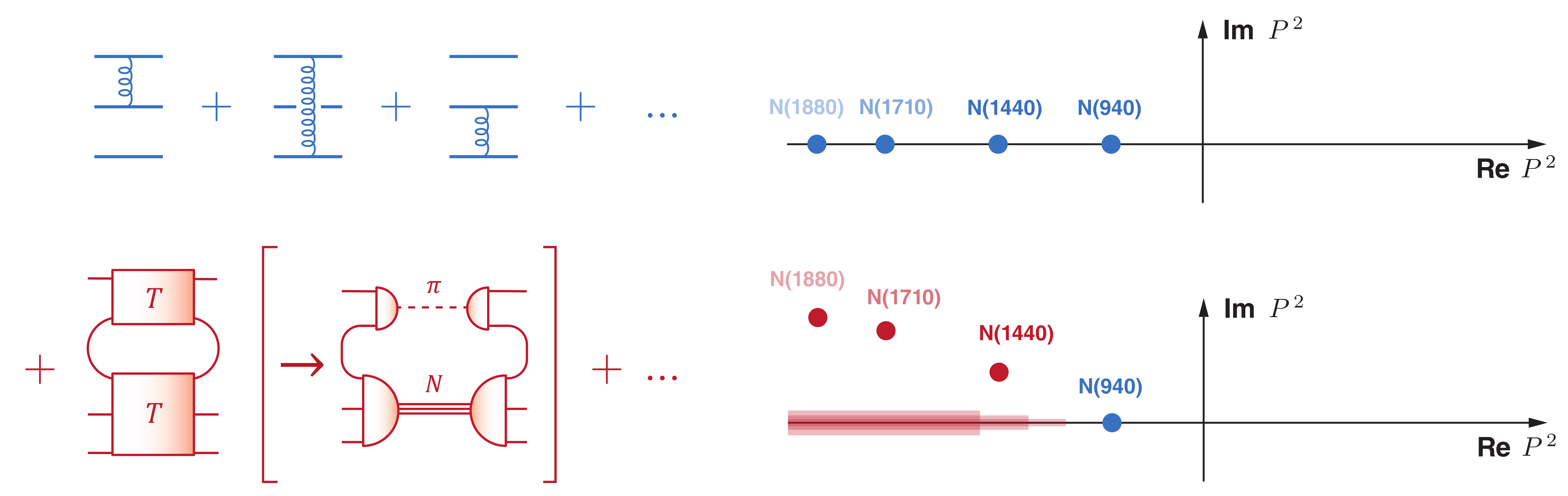}
              \caption{Rainbow-ladder like interactions produce real bound-state poles \textit{(top)}; to generate resonances,
                       a resonance mechanism in the Bethe-Salpeter kernel is needed \textit{(bottom)}.} \label{fig:resonances}
      \end{figure*}

     The generation of meson and baryon \textit{resonances} through BSEs is more intricate and sketched in Fig.~\ref{fig:resonances}.
     A gluon exchange between the quarks in the three-body BSE does not provide a decay mechanism and
     the resulting baryons are bound states on the real axis, where $P^2$ is the squared total Euclidean momentum.
     To produce multiparticle cuts and decay widths that push the poles into the complex plane, 
     one needs topologies such as those at the bottom of Fig.~\ref{fig:resonances}, where the
     $qqq$ and $q\bar{q}$ correlation functions can generate nucleons and pions as intermediate states inside the kernel.
     However, these can only enter beyond rainbow-ladder. 
     Similar kernels with intermediate $\pi\pi$ exchanges were recently included in the $\rho$-meson BSE and the quark-photon vertex,
     where they shift the $\rho$ pole into the complex plane and generate a width of the expected size~\cite{Williams:2018adr,Miramontes:2019mco}.
     
     The situation is different for tetraquarks obtained from a four-quark BSE, which is 
     the generalization of the BSE in Fig.~\ref{fig:nptfcts} to a $qq\bar{q}\bar{q}$ system.
     In that case the $q\bar{q}$ pairs in the equation can recombine to form meson poles,
     which dynamically generate decay thresholds and widths even in rainbow-ladder.
     The existing four-quark calculations reproduce the mass pattern of the light scalar mesons
     and the mass of the $X(3872)$~\cite{Eichmann:2015cra,Wallbott:2019dng}.

     \begin{figure}[t]
     \center{
     \includegraphics[width=1\textwidth]{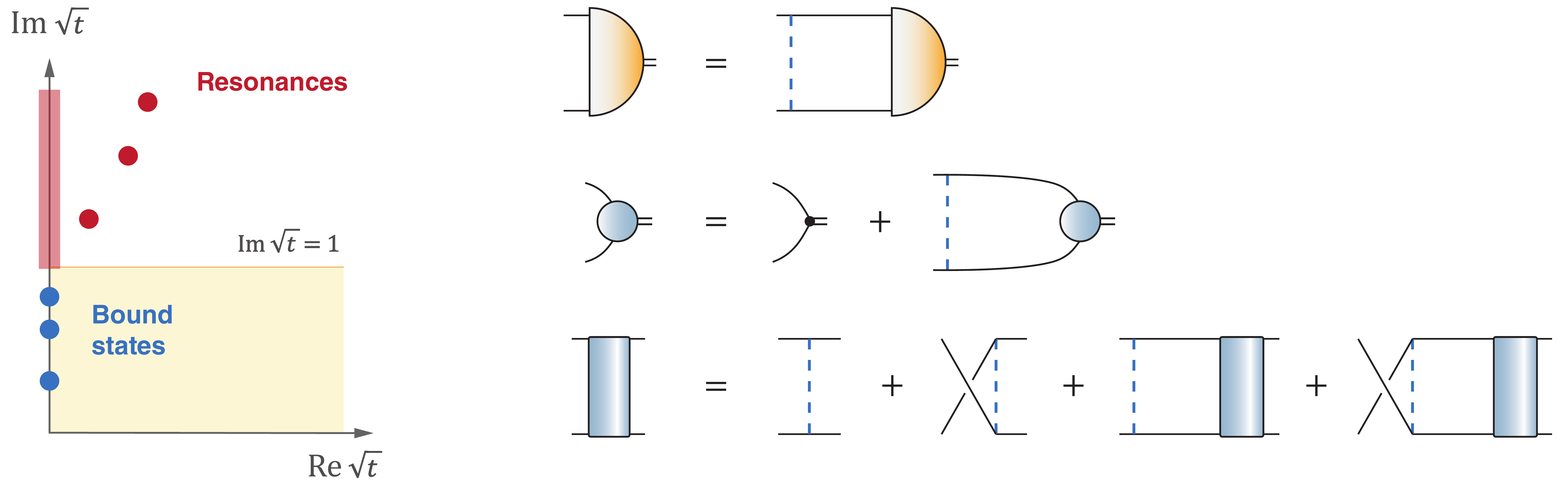}}
        \caption{\textit{Left:} Sketch of a typical singularity structure in the complex $\sqrt{t}$ plane. \textit{Right:} homogenous BSE, inhomogeneous BSE and scattering equation for a system with a ladder exchange.}
        \label{fig:eqs}
     \end{figure}
     
     \newpage
     
     In addition to the presence of a decay mechanism, 
     one must also build the necessary techniques to extract the resonance information. 
     Crossing a physical threshold means integrating over internal 
     ($\pi\pi$, $N\pi$, $\dots$) poles in a BSE. 
     In fact, already the quark and gluon correlation functions
     in the kernels contain singularities in the complex plane; those do not produce physical thresholds but still put limits on the calculable (`Euclidean')
     momentum regions.  
     As a consequence, one is restricted to low-lying meson and baryon excitations, states below thresholds, form factors within a certain $Q^2$ range, etc.
     For example, in Fig.~\ref{spectrum-light} the nearest complex singularities from the quark propagators in the BSE integrand put an 
     upper  limit on the masses of  excited states;
     to go further, one would need contour deformations analogous to those employed in Refs.~\cite{Williams:2018adr,Miramontes:2019mco,Maris:1995ns,Strauss:2012dg,Windisch:2012zd,Windisch:2012sz,Pawlowski:2017gxj,Weil:2017knt,Eichmann:2019dts}.

     In the following we summarize the contour-deformation technique in Ref.~\cite{Eichmann:2019dts} applied to a scalar BSE using a Euclidean metric. 
     This is equivalent to including the correct poles using residue calculus and
     thereby allows one to compare with Minkowski-space calculations~\cite{Kusaka:1995za,Sauli:2001we,Frederico:2013vga,Carbonell:2014dwa}.
     Consider two scalar particles with masses $m$ and $\mu$ (with $\beta=\mu/m$) and a three-point interaction with a dimensionless coupling $c=g^2/(4\pi m)^2$.
     The relevant equations are shown in Fig.~\ref{fig:eqs}: The first is the homogeneous BSE,
     which determines the mass spectrum and BS amplitude in a given $J^{PC}$ channel; the second is the inhomogeneous BSE
     which determines the corresponding vertex function; and the third is the scattering equation which determines
     the scattering amplitude. For simplicity we work with tree-level propagators and restrict the kernel to a scalar ladder exchange with mass $\mu$,
     which is the massive Wick-Cutkosky model~\cite{Wick:1954eu,Cutkosky:1954ru,Nakanishi:1969ph}.

     The left panel in Fig.~\ref{fig:eqs} shows the expected singularity structure in the complex plane of the variable $\sqrt{t} = \sqrt{P^2}/(2m)$. 
     The three equations are solved at fixed $t$.
     Because the tree-level propagators have real poles, the system has a threshold.
     Below the threshold $\text{Im}\sqrt{t}=1$ there can be bound states, whose locations depend on
     the parameters $c$ and $\beta$. The two-particle cut starts at the threshold and extends to infinity, and above the threshold one would expect
     resonances on the second Riemann sheet, with complex masses $M_i$ corresponding to $\text{Im}\sqrt{t} = \text{Re} \,M_i/(2m)$.

     To access the first sheet above the threshold, one must employ contour deformations:
     After integrating over the inner integration variable, the poles in the integrand from the constituent and exchange propagators
     become cuts and one has to deform the integration path in the outer integration variable to avoid those cuts.
     With contour deformations,
     the homogeneous BSE can be solved in the entire first sheet of the complex $\sqrt{t}$ plane.  

     Extracting the resonance locations requires access to the second sheet,
     which is not directly possible from numerical solutions of the (in-)homogeneous BSEs
     and relies on analytic continuations methods such as the Schlessinger-point method~\cite{Schlessinger:1968,Haritan:2017vvv,Tripolt:2017pzb}.
     One strategy is then to solve the BSE on the first sheet (below and above the threshold) and determine the singularity structure
     on the second sheet by analytic continuation.  
     The second option is to solve
     the scattering equation for the four-point function directly;
     from the resulting two-body unitarity relation one has direct access to
     the information on the second sheet, including the singularity structure of the amplitude.
     In turn, one has to deal with more complicated contour deformations
     and must first solve the half-offshell scattering equation from where the onshell
     scattering amplitude, where all four external legs are on their mass shells, is extracted in the end~\cite{Eichmann:2019dts}.

     \begin{figure}[t]
     \center{
     \includegraphics[width=1\textwidth]{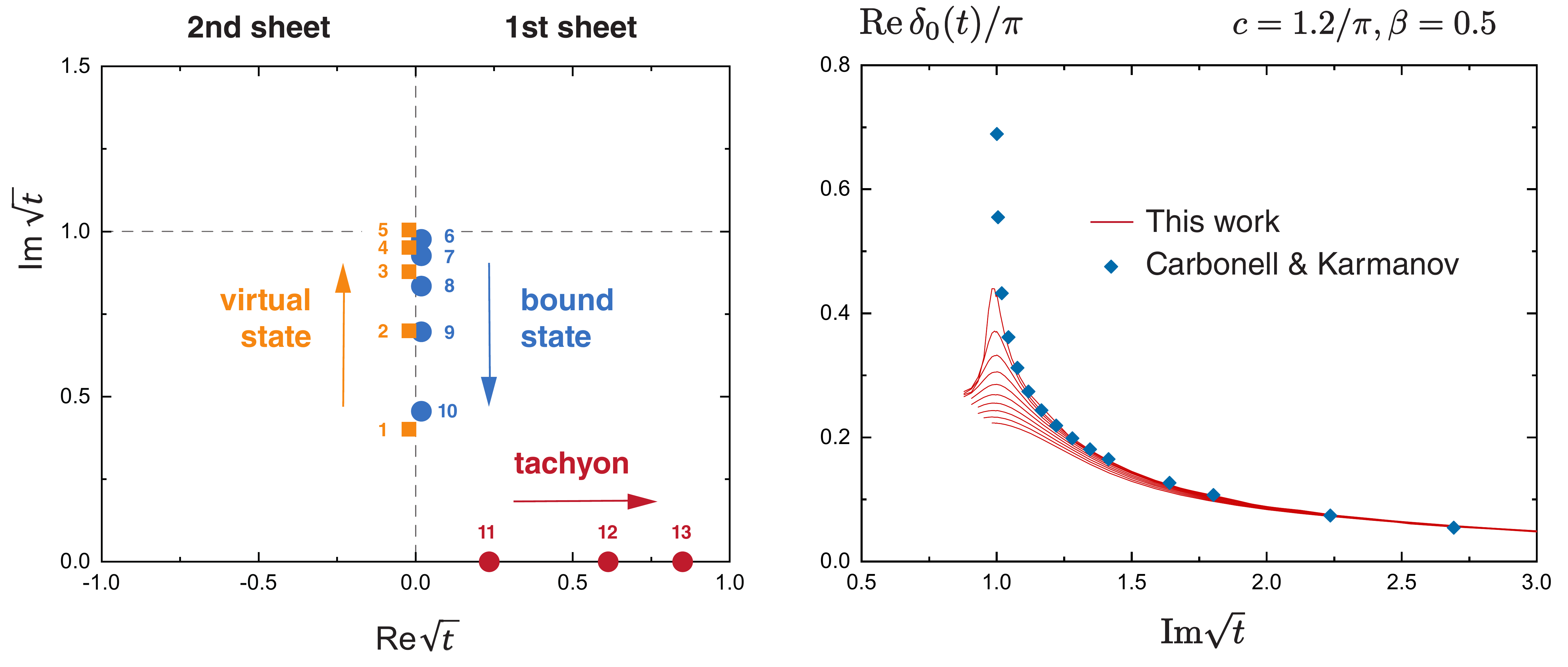}}
        \caption{\textit{Left:} Ground-state pole trajectory for $\beta=4$ as a function of the coupling~\cite{Eichmann:2019dts}.
        \textit{Right:} Phase shift for the leading partial wave compared to Ref.~\cite{Carbonell:2014dwa}.}
        \label{fig:results}
     \end{figure}

     The left panel in Fig.~\ref{fig:results} shows the resulting ground-state pole trajectory in the complex $\sqrt{t}$ plane
     determined from the scattering equation.
     It turns out that the scalar model does not produce resonances but instead virtual states on the imaginary $\sqrt{t}$ axis (or negative $t$ axis) of the second sheet.
     If one increases the coupling $c$ starting at $c=0$, the pole moves up to the threshold until it turns over to the first sheet, where the state becomes bound.
     When increasing the coupling further, the pole slides down on the first sheet until it eventually becomes tachyonic. The same pattern
     is repeated for excited states.

     The contour-deformation technique also allows one to extract phase shifts, which are related to the partial-wave amplitudes
     at $\text{Re}\sqrt{t}=0$ and $\text{Im}\sqrt{t} > 1$
     in a partial-wave expansion of the onshell scattering amplitude. The right panel in Fig.~\ref{fig:results} shows the resulting phase shift
     for the leading partial wave~\cite{bled}, which is a function of complex $\sqrt{t}$.
     We cannot extract the phase shifts directly \textit{on} the imaginary axis because here the deformed contour lies along a cut;
     instead we calculate it along lines in the complex $\sqrt{t}$ plane with $\text{Re}\sqrt{t}$ fixed and approaching the imaginary axis.
     The results are compared to those in Ref.~\cite{Carbonell:2014dwa}, where a Minkowski-space approach was employed to determine the phase shift
     for the same parameter set in the scalar model. Although moving closer to the axis requires increasingly better numerics, there is satisfactory agreement between the two approaches.

     In summary, contour deformations provide a practical toolkit for treating resonances with integral equations.
     The formalism can be taken over without changes to systems with spin, such as $N\pi$ or $NN$ scattering.
     Moreover, contour deformations are generally applicable for circumventing singularities in integrals and integral equations ---
     such as in Dyson-Schwinger and Bethe-Salpeter equations in QCD, where the singularities of the quark propagator and other correlation functions
     usually prohibit access to highly excited states, timelike form factors or form factors at large $Q^2$.

     \medskip

     \textbf{Acknowledgments:} I would like to thank the organizers of the Light Cone 2019 conference as well as my collaborators P.~Duarte, C.~S.~Fischer, M.~T.~Pe\~na, H.~Sanchis-Alepuz, A.~Stadler and P.~C.~Wallbott for their contributions summarized in this work.

\bibliographystyle{utphys-mod3}

\bibliography{lit-resonances}

\end{document}